\documentclass[a4paper,9pt]{article}
\usepackage[T1]{fontenc}
\usepackage[utf8]{inputenc}
\usepackage{lmodern}
\usepackage{graphicx}
\title{Hyperspherical three-body model calculation for the bound $^{1,3}$S-states of Coulombic systems}
\author{Md. Abdul Khan\\ 
\begin{footnotesize}
Department of Physics, Aliah University, 
\end{footnotesize}\\
\begin{footnotesize}
Action Area-II, Plot No. IIA/27, Newtown, Kolkata-700156, India.
\end{footnotesize}\\
\begin{footnotesize}
{\it Email:} drakhan.rsm.phys@gmail.com; drakhan.phys@aliah.ac.in
\end{footnotesize}
}
\vspace{1.0cm}

\begin{document}

\maketitle
\begin{abstract}
In this paper, hyperspherical three-body model formalism has been applied for the calculation energies of the low-lying bound $^{1,3}$S (L=0)-states of neutral helium and helium like Coulombic three-body systems having nuclear charge (Z) in the range Z=2 to Z=92. The calculation of the coupling potential matrix elements of the two-body potentials has been simplified by the introduction of Raynal-Revai Coefficients (RRC). The three-body wave function in the Schr\H{o}dinger equation when expanded in terms of hyperpherical harmonics (HH), leads to an infinite set of coupled differential equation (CDE). For practical reason the infinite set of CDE is truncated to a finite set and are solved by an exact numerical method known as renormalized Numerov method (RNM) to get the energy solution (E). The calculated energy is compared with the ones of the literature.
\end{abstract}
\hspace{1.0cm}{\it Keywords:} Raynal Revai Coefficient, Hyperspherical Harmonics, \\ \hspace*{1.0cm}Coupled Differential Equation, Potential Matrix Elements,\\ \hspace*{1.0cm}Renormalized Numerov Method.\\ \hspace*{1.0cm}{\it PACS:} 02.70.-c, 31.15.-Ar, 
31.15.{\cal J}a, 36.10.{\cal E}e.

\section{Introduction}
In physics, role of few-body (two- or three-body) problems are very important for the proper understanding of the physics underlying the internal configurations and kinematics of more complex many-body systems, usually made of interacting bosons and (or) fermions. As few-body systems are the building blocks of more complex many-body systems, they are important not only in nuclear, particle, plasma, astro-nuclear or hyper-nuclear physics but in atomic physics as well. For example, the lightest few-body systems like neutral helium atom and helium-like ions have long history as subject of attraction for both theoretical and experimental investigations. These atomic systems constituting the simplest few-body problems in atomic physics are traditionally used as testing ground for different methods of description of the structure of atoms. On the experimental side, small natural line widths of transition among various metastable quantum states of helium-like systems allow spectroscopic measurements of very high precision. In addition, few-body systems made up of electrons, protons, muons, deuteron, kaon  etc. and their antimatters are found to be of strong interest in many areas of physics including atomic spectroscopy, quantum electrodynamics, particle physics and astrophysics [1-2]. In recent years, highly ionized atoms are being studied extensively to explain the origin of X-rays spectra from the solar corona and other astrophysical plasmas. It is worth mentioning here that highly ionized atoms can be produced in the laboratory by collision of ions with atoms or directing energetic projectile beams towards matter foils and their spectra can also be studied in the laboratory.

A number of theoretical methods have been adopted to investigate the bound state properties of atomic few-body systems. For example, we may refer the works of Lin  [3-5], Lin et al [6] in which the author(s) has (have) applied hyperspherical coordinates to Coulombic three-body systems to calculate channel potential, channel function, binding energies and some other observables of the systems. Huang [7] investigated muonic helium atom as a three-body problem in correlated wave function approach. Some more works which may be referred here include those found in references [8-17]. Rajaraman et al [18] presented results of three-body problems originated in the nuclear matter. Alexander et al [19] reported  analytical results for the trimmer binding energies and other three-body parameters considering three-body system of identical bosonic atoms. Frolov [20] adopted exponential expansion based variational approach to construct highly accurate wave functions for the triplet spin states of helium like two-electron ions like Li$^+$ (atomic number Z=3] to Ne$^{8+}$ (atomic number Z=10]. The ground state properties of some two-electron and electron-muon atomic three-body systems has been studied by Rodriguez et al [21] applying the Angular Correlated Configuration Interaction (ACCI) approach. The calculated energy for negatively charged hydrogen-like systems; neutral helium-like systems, and positively charged lithium like systems. However, more accurate results for these systems are reported by Smith Jr et al [22], Frolov et al [23-31], Thakkar and Koga [32], Goldman [33], Korobov [34] and Drake [35]. Researchers like- Hylleraas and Ore [36], Hill [37], Mohr and Taylor [38], Drake, Cassar and Razvan [39], Frolov [40], Mills [41-42], Ho [43-44] and Wen-Fang [45] have explored the bound state properties of exotic positronium negative ion $Ps^- (e^+e^-e^-)$. Ancarani et al [1-2] also reported the ground and excited state energies for several three-body atomic systems obtained by applying ACCI approach. Kubicek et al [46] and  Kondrashev et al [47] conducted experiments on production of He-like ions.
In this paper we present energies of the low-lying bound $^{1,3}$S-states of neutral helium and helium like two-electron Coulombic three-body systems having nuclear charge number Z in the range Z=2 - 92. The resulting three-body Schr\H{o}odinger equation has been solved in the framework of hyperspherical harmonics expansion (HHE) formalism applying an exact numerical method known as renormalized Numerov method (RNM) [48]. 
The scheme of solution of the three-body Schr\H{o}dinger equation in HHE approach has been described in more details in our earlier works [49-62].
 In HHE approach for a general three-body system containing particles of arbitrary masses, there are three possible partitions and in the $i^{th}$ partition, the particle labeled $i$, acts as a spectator while the remaining two,  labeled $j$ and $k$ form the interacting pair. For the calculation matrix element of the potential of the $(jk)$ pair, V($r_{jk}$), it is convenient to expand the chosen HH in the set of HH corresponding to the partition in which the potential $\vec{r_{jk}}$ is proportional to the first Jacobi vector $\vec{\xi_i}$ [49] and this has been done using Raynal-Revai coefficients (RRC) [63]. In the numerical procedure of computation of potential matrix elements of the two-body potentials involved in the system of three particles constituted by a relatively heavy and positively charged nucleus being orbited by two valence electrons,  we used RRC of [49,64]. The energies of the low-lying bound $^{1,3}$S-states of several Coulombic three-body systems obtained by solving the three-body Schr\"{o}dinger equation have been compared with the ones of the literature. 

In Section 2, we will give a very concise description of the HHE method along with the scheme of transformation between two sets of HH which correspond to two different partitions. In Section 3, we will discuss the application of HHE to the low-lying bound spin singlet (spin S=0) and spin triplet (spin S=1) i.e. $^{1,3}$S (L=0)-states of neutral helium and similar other systems to calculate the energies and compare them with the ones of the literature.

\section{HHE Method}
In the hyperspherical harmonics expansion (HHE) method for a general three-body system of particles of arbitrary masses $m_{i}$, $m_{j}$, $m_{k}$ as depicted in Figure 1,  
\begin{figure}
\centering
\fbox{\includegraphics[width=0.95\linewidth, height=0.75\linewidth]{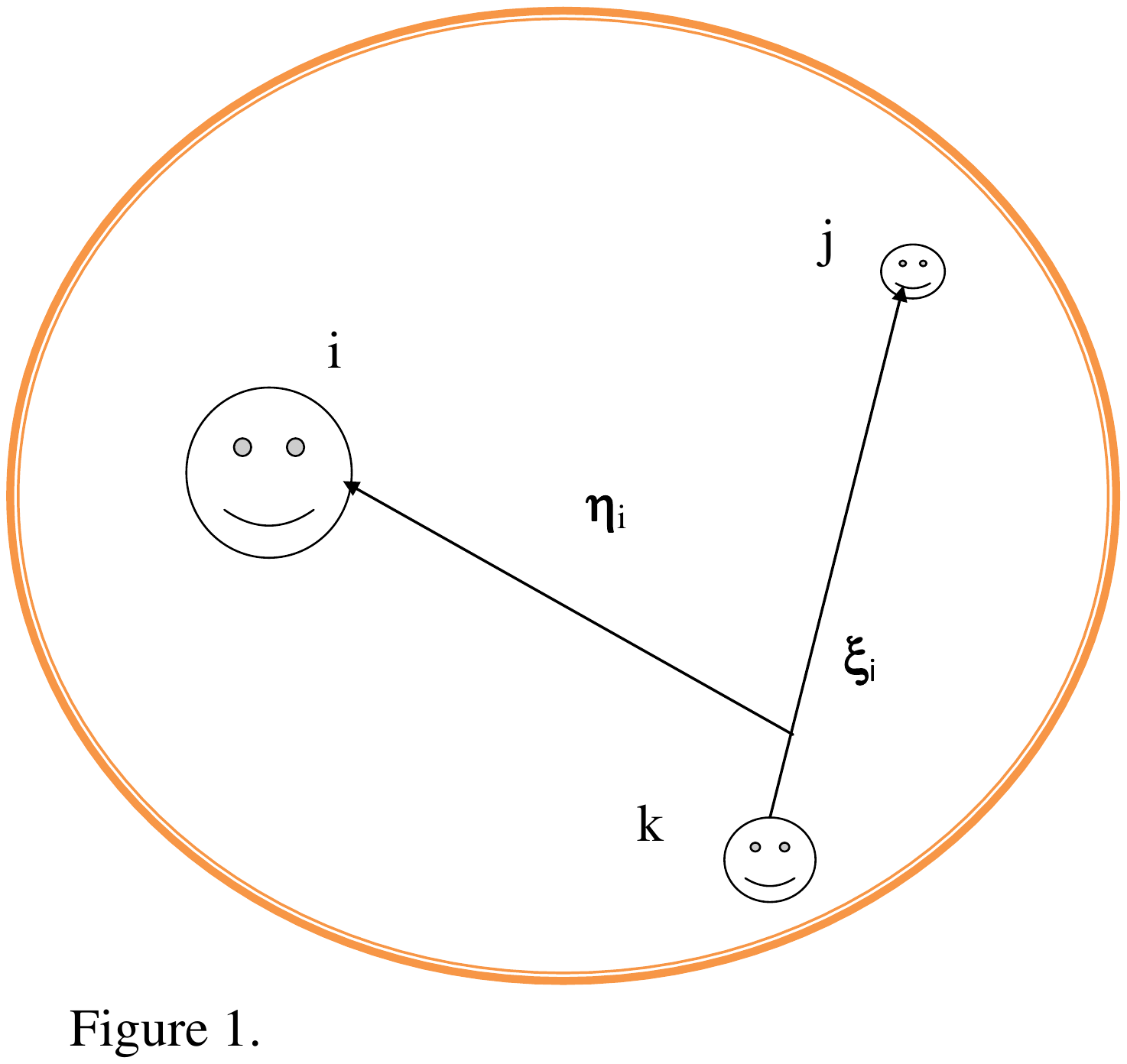}}
\caption{Particle label scheme for general three-body system and choice of Jacobi coordinates in the $i^{th}$ partition.}
\label{fig:boxed_graphic}
\end{figure}
the Jacobi coordinates [64] in the partition - $\lq\lq i$" are defined as 
\begin{equation}
 \left.  \begin{array}{ccl}
   \vec{\xi_{i}} & = & \left[ \frac{m_{j} m_{k}M}{m_{i}(m_{j}+m_{k})^{2}}
\right] 
^{\frac{1}{4}} (\vec{r_{j}} - \vec{r_{k}}) \\ 
   \vec{\eta_{i}} & = & \left[ \frac{m_{i} (m_{j}+m_{k})^{2}}
{m_{j} m_{k} M}\right]^{\frac{1}{4}} \left( \vec{r_{i}} - \frac{m_{j}
\vec{r_{j}} + m_{k} \vec{r_{k}}}{ m_{j} + m_{k}} \right) 
    \end{array}  \right\} 
\end{equation}
where $M=m_i+m_j+m_k$ and the condition that ($ i, j, k $) should form a cyclic permutation of (1, 2, 3) determines the sign of $\vec{\xi_{i}}$.\\
The set of Jacobi coordinates represented by eq.(1) above corresponds to the partition, in 
 which, the particle labeled $\lq\lq i$" is the spectator and the remaining particles labeled  $\lq\lq j$" and $\lq\lq k$" form the interacting pair. The reason behind such nomenclature is that the calculation of matrix element of V$(\vec{r}_{jk})$ in terms of the above set of Jacobi coordinates is straight forward. In the similar manner, we can also define two other sets of Jacobi coordinates by cyclically permuting $i\rightarrow j\rightarrow k\rightarrow i$ twice, which correspond to $j^{th}$ and $k^{th}$ partitions respectively.\\
In hyper-spherical variables [65-66] of the $i^{th}$ partition, three-body Schr\H{o}dinger equation is  
\begin{equation}
\left[ - \frac{\hbar^2}{2\mu\rho^5}
\frac{\partial}{\partial\rho}(\rho^5 \frac{\partial}{\partial\rho}) + 
\frac{\hbar^2}{2\mu\rho^2}\frac{\hat{{\cal N}}^{2}(\Omega_{i})}{\rho^{2}} + V (\rho, \Omega_{i}
) - E \right] 
\Psi (\rho, \Omega_{i} ) \:=\: 0
\end{equation}
where ${\mu\: =\: \left[ \frac{m_{i} m_{j} m_{k}}{M} \right]^{\frac{1}{2}}}$ is an effective
 mass parameter, $V (\rho, \Omega_{i})$ = $ V_{jk} + V_{ki} + V_{ij} $ is the
total  interaction potential, and $\hat{\cal N}^{2}(\Omega_{i})$ is the square
of the hyper angular momentum operator satisfying the eigenvalue equation [67]
\begin{equation}
\hat{{\cal N}}^{2}(\Omega_{i}) {\cal H}_{N \alpha_{i}}(\Omega_{i}) \:=\: N
( N + 4 )  
{\cal H}_{N \alpha_{i}}(\Omega_{i})
\end{equation}
where 
\begin{equation}
 \begin{array}{rcl}
{\cal H}_{N \alpha_{i}}( \Omega_{i} )& \equiv &
{\cal H}_{N l_{\xi_{i}} l_{\eta_{i}} L M}(\phi_{i}, \theta_{\xi_{i}},
\phi_{\xi_{i}}, \theta_{\eta_{i}}, \phi_{\eta_{i}})\\
 & \equiv & ^{(2)}P_{N}^{l_{\xi_{i}} l_{\eta_{i}}}(\phi_{i})
\left[ Y_{l_{\xi_{i}} m_{\xi_{i}}}(\theta_{\xi_{i}}, \phi_{\xi_{i}}) Y_{l_{\eta_{i}}
m_{\eta_{i}}}(\theta_{\eta_{i}}, \phi_{\eta_{i}}) \right]_{L M}
 \end{array}
\end{equation}
is the normalized eigenfunction known as hyperspherical harmonics (HH),  $L$ is the total orbital angular momentum of the system with $M$ as its projection, 
$\Omega_{i} \rightarrow \{ \phi_{i}, \theta_{\xi_{i}}, \phi_{\xi_{i}},
\theta_{\eta_{i}}, \phi_{\eta_{i}} \}$,  ${ \alpha_{i} \: \equiv \: \{ l_{\xi_{i}}, l_{\eta_{i}}, L, M \} }$  is
a short hand notation and $[ ]_{L M}$ indicates angular momentum coupling.
The quantity $N=2n_i+ l_{\xi_i} + l_{\eta_i}$ ($n_i$ being a non-negative 
integer) is the hyper-angular momentum quantum number which is not a good
quantum number for the three-body system. 
In terms of HH associated with a given partition, (say partition $\lq\lq i$), the wave-function $\Psi (\rho, \Omega_{i})$ is expanded in the complete set of HH 
\begin{equation}
\Psi(\rho, \Omega_{i}) = \sum_{N\alpha_{i}}\frac{U_{N\alpha_{i}}   
  (\rho)}{\rho^{5/2}} {\cal H}_{N\alpha_{i}}(\Omega_{i})
\end{equation}
 Substitution of eq.(5) in eq.(2),  use of eq.(3) and the ortho-normality of HH, leads to a set of coupled differential equations (CDE) in $\rho$
\begin{equation}
\begin{array}{cl}
& \left[ -\frac{\hbar^{2}}{2\mu} \frac{d^2}{d\rho^2}
+\frac{\hbar^{2}}{2\mu}\frac{(N+3/2)(N+5/2)}{\rho^2} - E \right] 
U_{N\alpha_{i}}(\rho)  \\
+ & \sum_{N^{\prime} \alpha_{i}^{~\prime}} <N\alpha_{i}
\mid V(\rho, \Omega_{i}) \mid N^{\prime} \alpha_{i}^{~\prime}
>U_{N^{\prime} \alpha_{i}^{~\prime}}(\rho) \: = \: 0.
\end{array}
\end{equation}
where
\begin{equation}
<N \alpha_{i} |V| N^{\prime}, \alpha_{i}^{\prime}> = \int
{\cal H}_{N\alpha_{i}}^{*}(\Omega_{i}) V(\rho, \Omega_{i}) {\cal
H}_{N^{\prime} 
 \alpha_{i}^{~\prime}}(\Omega_{i}) d\Omega_{i}
\end{equation}
For central potentials, computation of the matrix elements of the form 
$$<{\cal H}_{N \alpha_{i}}(\Omega_{i}) \mid V_{j k}(\xi_{i}) \mid {\cal
H}_{N^{\prime}  \alpha_{i}^{\prime}}(\Omega_{i})>$$ is
straight forward, while for matrix elements of the forms $$<{\cal H}_{N
\alpha_{i}}(\Omega_{i}) \mid V_{k i} (\xi_{j})|{\cal H}_{N^{\prime}
 \alpha_{i}^{~\prime}}(\Omega_{i})>$$ or $$<{\cal H}_{N
\alpha_{i}}(\Omega_{i}) \mid V_{i j}(\xi_{k}) \mid {\cal H}_{N^{\prime}
 \alpha_{i}^{~\prime}}(\Omega_{i})>$$ computations become very
complicated even for central potentials. This is because the vectors $\vec{\xi_j}$ or $\vec{\xi_k}$ depend 
on the polar angles of the vectors $\vec{\xi_{i}}$ and $\vec{\eta_{i}}$. Vectors $\vec{\xi_{k}}$ and $\vec{\eta_{k}}$ can be expressed in terms of $\vec{\xi_{i}}$ and $\vec{\eta_{i}}$ using eq.(1) as
\begin{equation}
 \left. \begin{array}{rcl}
\vec{\xi_{k}} & = & - \cos \sigma_{ki} \vec{\xi_{i}} + \sin \sigma_{ki}
\vec{\eta_{i}}\\ 
\vec{\eta_{k}} & = & - \sin \sigma_{ki} \vec{\xi_{i}} - \cos \sigma_{ki} \vec{\eta_{i}}
	  \end{array} \right\}
\end{equation}
where $\sigma_{ki}$ = $\tan^{-1} \{(-1)^{P} \sqrt{\frac{M m_{j}}{m_{i} m_{k}}}
\}$, P being even (odd) if ($kij$) is an even (odd) permutation of the
triad (1 2 3).
Now for any arbitrary shape of the central potential with non-vanishing L, 
most of the five dimensional integrals have to be done numerically which 
makes the calculation time consuming and inaccurate. However computation 
of the latter matrix elements can be greatly simplified using the following
prescription. At first it is to be noted that each of the complete sets of HH 
$\{{\cal H}_{N \alpha_i}(\Omega_i)\}$, $\{{\cal H}_{N
\alpha_{j}}(\Omega_{j})\}$ or $\{{\cal H}_{N \alpha_k}(\Omega_k)\}$ span 
the same five dimensional angular hyperspace. Then a particular member of a given set, say
${\cal H}_{N \alpha_i}(\Omega_i)$ can be expanded in the complete set of
$\{{\cal H}_{N \alpha_k}(\Omega_k)\}$ through a unitary transformation:
\begin{equation}
{\cal H}_{N \alpha_i}(\Omega_i)=\sum_{\alpha_k} <\alpha_k\mid
 \alpha_i>_{NL} {\cal H}_{N\alpha_k}(\Omega_k) 
\end{equation}
As $N, L, M$ are conserved for eq.(9) and there is rotational degeneracy with 
respect to the quantum number $M$ for spin independent forces, we have 
\begin{equation}
<\alpha_{k} \mid \alpha_{i}>_{NL} = <l_{\xi_{k}} l_{\eta_{k}}\mid l_{\xi_{i}}
l_{\eta_{i}}>_{NL}
\end{equation}
Thus, eq.(9) can be rewritten as [52]
\begin{equation}
{\cal H}_{N\alpha_{i }}(\Omega_{i}) = \sum_{l_{\xi_k} l_{\eta_k}}<l_{\xi_k}
l_{\eta_k}\mid l_{\xi_i} l_{\eta_i}>_{NL} {\cal H}_{N\alpha_k}(\Omega_k)
\end{equation}
 The coefficients involved in eq.(10) and (11) are called the Raynal-Revai Coefficients (RRC) and these are independent of M due to overall rotational degeneracy. In terms of these coefficients, the 
matrix element of a central interaction $V_{ij}$ then becomes 
\begin{equation}
 \begin{array}{ll}
&<{\cal H}_{N\alpha_i}(\Omega_i) \mid V_{ij} (\xi_k)\mid {\cal H}_{N^{\prime} 
 \alpha_i^{\prime}}(\Omega_i)>\\
=&\sum_{l_{\xi_k}^{\prime}
l_{\eta_k}^{\prime} l_{\xi_k} l_{\eta_k}}<l_{\xi_k} 
l_{\eta_k} \mid l_{\xi_{i}} l_{\eta_{i}}>_{NL}^{*}
\times<l_{\xi_k}^{\prime}
l_{\eta_k}^{\prime}\mid l_{\xi_i}^{\prime} l_{\eta_i}^{\prime}>_{N^{\prime}L}\\ 
\times & <{\cal H}_{N\alpha_k}(\Omega_k)\mid V_{ij}(\xi_k)\mid {\cal H}_{N^{\prime}
 \alpha_k^{\prime}}(\Omega_k)>
 \end{array}
\end{equation}
 The matrix element on the right side of eq.(12) has the same form as the
matrix element of $V_{jk}$ in the partition $i$ and can be calculated in a
straight forward manner. Thus computing the values of RRC's involved in eq.(12) 
 using their explicit expressions found in [49, 63], one can calculate the matrix 
element of $V_{ij}$ easily. Similar prescription can also be
employed for the calculation of the matrix element of $V_{ki}$.

\section {Application to Coulombic three-body systems}
\hspace*{1cm} We apply the scheme of RRC to the low-lying bound $^{1,3}$S (L=0)-states of Coulombic three-body system containing relatively massive and positively charged nuclear core plus two extra core orbital electrons. We label the nuclear core having mass $m_C$ and charge +Ze as the $i^{th}$ particle, two electrons of mass $m_j =m_k = m$  and charge -e as the $j^{th}$ and $k^{th}$ particles respectively. For this particular choice mass of the system particles, Jacobi coordinates of eq.(1) in corresponding to the partition $ i$ becomes 
\begin{equation}
 \left. \begin{array}{rcl}
  \vec{\xi_{i}} & = &\beta_{i} (\vec{r_{j}} - \vec{r_{k}}) \\
  \vec{\eta_{i}} & = &\frac{1}{\beta_{i}} (r_{i} - \frac{\vec{r_{j}}+
\vec{r_{k}}}{2}) 
	  \end{array} \right\}
\end{equation}
where the dimensionless parameter $\beta_{i}$ = $\left[ \frac{m_{C}+2 m}{4 m_{C}} \right]^{\frac{1} {4}}$ can be connected to the effective mass $\mu$ as  
\begin{equation}
 \begin{array}{cclcl}
\mu & = & m \left( \frac{m_{C}}{m_{C}+2m} \right)^{\frac{1}{2}} & = &
\frac{m}{2\beta_{i}^{2}} 
 \end{array} 
\end{equation}
In atomic unit (ie., $\hbar^{2}$=$m$=$e^{2}$=1), eq.(6) becomes 
\begin{equation}
\left. \begin{array}{lcl}
 \left[-\beta_i^2 \frac{d^{2}}{d\rho^{2}}
+\beta_i^2 \frac{(N+3/2)(N+5/2)}{\rho^{2}}-E \right] 
U_{N\alpha_{i}}(\rho)&&  \\
+ \sum_{N^{\prime} \alpha_{i}^{\prime}} <N\alpha_{i}
\mid \frac{\beta_i}{\rho 
cos \phi_i} - \frac{Z}{\rho\left|(\beta_i sin \phi_{i})
\hat{\eta_i}- (\frac{1}{2\beta_i} cos\phi_i)\hat{\xi_i}\right|}&& \\
  -\frac{Z}{\rho\left|
(\beta_i sin \phi_i) \hat{\eta_i}+(\frac{1}{2\beta_i}cos\phi_i)
\hat{\xi_i}\right|}\mid N^{\prime}\alpha_i^{\prime}
>U_{N^{\prime}\alpha_i^{\prime}}(\rho)& = & 0
\end{array} \right\}
\end{equation}
 Mass of the particles involved in the present calculation is partly taken from [1-2, 28-30, 68-69]. A straight forward evaluation of the matrix 
elements of last two terms in eq.(15) would be prohibitively involved both for analytical
reduction to a computationally feasible form, as well as for the numerical
calculation. Furthermore, the numerical calculation would be both inaccurate and time
consuming. Application of RRC greatly simplifies the calculation, since in the 
partitions $k$ and $j$,  the third and
fourth terms inside ket-bra <> notation in eq.(15) is reduced to $\frac{Z\beta_{k}}{\rho cos \phi_{k}}$ and  $\frac{Z\beta_{j}}{\rho cos \phi_{j}}$ respectively. In the case of two-electron ions,
\begin{equation}
 \begin{array}{ccccl} \beta_{j}& = &\beta_{k}& = &\left[1-\frac{m^{2}}{(m_{C} +m)^{2}}
\right]^{\frac{1}{4}}  \end{array} \end{equation}
In the case of a heavy nucleus, ${m_{C}\gg m}$ and ${\beta_{i} \approx
\frac{1}{\sqrt{2}}}$, ${\beta_{j}=\beta_{k} \simeq}$ 1.\\ 

We expand the three-body relative wave function in the complete set of HH
 appropriate to the partition $i$ according to eq.(5). For the low-lying 
$^{1,3}$S-states of two-electron systems the total orbital angular momentum, $L$=0. 
Consequently ${l_{\xi_{i}}= l_{\eta_{i}}}$. Hence the set of quantum numbers represented 
by $\alpha_{i}$ is ${\left\{ l_{\xi_{i}}, l_{\xi_{i}}, 0, 0 \right\}}$ and the quantum numbers
${\left\{N\alpha_{i} \right\}}$ can be represented by ${\left\{N l_{\xi_{i}}\right\}}$ only. Furthermore for S=0, spin part of the two-electron wave function is anti-symmetric, hence the space part of the wave function must be symmetric under exchange of two electrons which allows only even values of  $l_{\xi_{i}}$ ($\leq N/2$). On the other hand for S=1, spin part of the wave function is symmetric, hence space part of the wave function must be anti-symmetric under the exchange of the two electrons which allows only odd values of  $l_{\xi_{i}}$ ($\leq N/2$) are to be considered. Corresponding HH is then given by [65]
\begin{equation}
\left. \begin{array}{ccl}
{\cal H}_{N \alpha_{i}}(\Omega_{i}) & \equiv & {\cal H}_{N l_{\xi_{i}}
l_{\xi_{i}} 0 0}(\Omega_{i}) \\ 
 & =  & ^{(2)}P_{N}^{l_{\xi_{i}} l_{\xi_{i}}}(\phi_{i})\left[Y_{l_{\xi_{i}}
m_{\xi_{i}}}(\theta_{\xi_{i}},\phi_{\xi_{i}}) 
Y_{l_{\xi_{i}} - m_{\xi_{i}}} (\theta_{\xi_{i}},\phi_{\xi_{i}})\right]_{0 0} \\
 &  &N \: even \: and \:  l_{\xi_{i}}=0, 2, 4, \ldots , \leq N/2 \:for\: ^1S\: states; \\
&  & l_{\xi_{i}}=1, 3, 5, \ldots , \leq N/2 \:for\: ^3S\: states
 \end{array}	\right\}
\end{equation}
 The matrix element of the two electron repulsion term in our chosen partition
$i$ is 
\begin{equation}
  \begin{array} {rcl}
    <N^{\prime}l_{\xi_{i}}^{\prime}|\frac{\beta_{i}}{\rho cos\phi_{i}}|N
l_{\xi_{i}}>&=&\frac{\beta_{i}}{\rho} \delta_{l_{\xi_{i}}^{\prime}, 
 l_{\xi_{i}}} \int_{0}^{\pi /2} { ^{(2)}P_{K^{\prime}}}^{l_{\xi_{i}}
l_{\xi_{i}}}(\phi) \\
&&\times {^{(2)}P_{N}}^{l_{\xi_{i}} l_{\xi_{i}}}(\phi) \sin^{2}\phi \cos\phi d\phi\\
  \end{array}
\end{equation}
in which suffix $i$ on $\phi$ has been dropped deliberately as it is only a
variable of integration. Similarly the matrix element of the third term of the total potential in eq.(15) in the partition $k$  is 
\begin{equation}
  \begin{array} {rcl}
<N^{\prime} l_{\xi_{k}}^{\prime} \mid \frac{\beta_{k}}{\rho cos\phi_{k}}
\mid N l_{\xi_{k}}>&=& \frac{\beta_{k}}{\rho} \delta_{l_{\xi_{k}}^{\prime},
l_{\xi_{k}}} \int_{0}^{\pi /2}  {^{(2)}P_{N^{\prime}}}^{l_{\xi_{k}}
l_{\xi_{k}}}(\phi) \\
&&\times {^{(2)}P_{N}}^{l_{\xi_{k}} l_{\xi_{k}}}(\phi) \sin^{2}\phi \cos~\phi ~d\phi\\
  \end{array}
\end{equation}
 A similar relation holds for the matrix element of the last term of the total potential in eq. (15)  in
the partition $j$. Eq.(18) and (19) show that the matrix elements are
essentially the same in the respective partitions, although $l_{\xi_{k}}$
and $l_{\xi_{j}}$ are not restricted to only even or odd integer values. Each 
 involves only a single, one dimensional integral to be performed
numerically. Using eq.(12), matrix elements of the third and fourth terms
of the total potential in eq.(15) in the partition $i$ become
\begin{equation}
 \begin{array}{rcl}
<N^{\prime} l_{\xi_{i}}^{\prime}\mid\frac{Z}{r_{ij}}\mid N
l_{\xi_{i}}> & = & \sum_{l_{\xi_{k}}}<l_{\xi_{k}}l_{\xi_{k}} \mid
l_{\xi_{i}}^{\prime}l_{\xi_{i}}^{\prime}>_{K^{\prime}0}^{\*} 
<l_{\xi_{k}} l_{\xi_{k}}\mid l_{\xi_{i}} l_{\xi_{i}}>_{N 
0} \\
 && <K^{\prime} l_{\xi_{k}}\mid\frac{Z\beta_{k}}{\rho
cos\phi_{k}}\mid K l_{\xi_{k}}>. 
 \end{array}
\end{equation}
and
\begin{equation}
 \begin{array}{rcl}
<N^{\prime}l_{\xi_{i}}^{\prime}\mid\frac{Z}{r_{ik}}\mid N l_{\xi_{i}}>
&=&\sum_{l_{\xi_{j}}} <l_{\xi_{j}}l_{\xi_{j}}\mid 
 l_{\xi_{i}}^{~\prime}l_{\xi_{i}}^{\prime}>_{N^{\prime}0}^{\*}
<l_{\xi_{j}}l_{\xi_{j}}\mid 
l_{\xi_{i}}l_{\xi_{i}}>_{N0} \\ 
 &&<N^{\prime}l_{\xi_{j}}\mid\frac{Z\beta_{j}}{\rho
cos\phi_{j}}\mid Nl_{\xi_{j}}>.  
 \end{array}
\end{equation}
In eq.(20) and (21) sums over $l_{\xi_{k}}^{\prime}$ and
$l_{\xi_{j}}^{\prime}$  respectively have been performed using the Kronecker
 - $\delta$'s in eq.(19) and a similar one with suffix $k$ replaced by suffix
 $j$. Thus the calculation of the matrix elements of $\underline{all}$ the
interactions become practically simple and easy to handle numerically.

Although the rate of convergence of HH expansion is reasonably fast [70] for short-range interaction potentials [67, 70-71],  the same cannot be claimed for long-range coulomb potentials. So to achieve desired convergence, large enough $N_m$ value is to be included in the calculation. In the case of singlet spin (S=0) states, if all N values up to a maximum of $N_m$ are retained in the HH expansion then the number of such basis function is 
\begin{equation}
 N_B=\left\{ \begin{array}{l}
  (\frac{N_m}{4}+1)^2 \: \:if \: \frac{N_m}{2} \: is\: even\\
  \frac{(\frac{N_m}{2}+1)(\frac{N_m}{2}+3)}{4} \: \:if\: \frac{N_m}{2} \: is \: odd. 
	  \end{array} \right.
\end{equation}
and that for the triplet spin state (S=1) is 
\begin{equation}
 N_B=\left\{ \begin{array}{l}
  (\frac{N_m}{4})(\frac{N_m}{4}+1) \: \:if \: \frac{N_m}{2} \: is\: even\\
  (\frac{N_m+2}{4})^2 \:if\: \frac{N_m}{2} \: is \: odd. 
	  \end{array} \right.
\end{equation}
It can be checked from eq.(22) and eq. (23) respectively, that the number of basis states ($N_B$) and hence the size of CDE \{eq.(6)\} increases rapidly as $N_m$ increases. For example, for $N_m=96$ one has to solve 625 CDE for singlet spin and 600 CDE for the triplet spin configuration respectively which lead the calculation towards instability. 
We used dual-core based desktop computer for the present calculation and could solve only up to $N_m=28$ reliably. The calculated binding energy ($B_{N_m}$) for values of $N_m$ up to 28 are presented in columns 2 -10 of Table I for few low-lying $^{1,3}$S-states of two-electron Coulombic systems like neutral helium and highly ionized radon Rn$^{84+}$ and that for Rb$^{35+}$ is presented in Table 3. The energies for still higher $N_m(>28)$ may be obtained by following an extrapolation theorem suggested by Schneider [72] as discussed below. According to the theorem on convergence of HH one may expect following relation to hold for coulomb interaction:
\begin{equation} (N_m+y)^4\Delta B_{N_m}=C, \end{equation}
where \begin{equation} \Delta B_{N_m}=E_{N_m+4}-B_{N_m} \end{equation} and y, C are
constants. If one obtains y and C by solving eq.(24) for $N_m$=16 and 20 
and uses eq.(24) to estimate the BE for $N_m=28$, he finds that the estimated 
BE agrees fairly well with the BE actually calculated by solving CDE with $N_m=28.$ 
In this way one may verify that eq.(24) is well obeyed. For the converged 
extrapolated BE (=$B_{con}=B_{N_M}$), we calculated the constants y and C 
from eq.(24) by least square fitting $B_{N_m}$ obtained by solving the 
CDE for $N_m=0, 4, 8, ..., up\; to\; 28$. With these values of y and C 
the extrapolated energies calculated by eq.(24) for larger $N_m$ (>28) are 
presented in columns 2-10 of Table 2. We select a value of $N_m (=N_M)$ for which $\Delta B_{N_m}$ is of the same order as the overall numerical error ($\epsilon$) in the calculation of $B_{N_m}$ with $N_m\leq28$. Since, for 
any $N_m>N_M$, the correction $\Delta B_{N_M}$ will be smaller than $\epsilon$ and hence unreliable due to the finite numerical precision in solving the CDE. We estimated $\epsilon$ to be about $4\times10^{-5}$ for a double precision calculation using dual-core based personal computer. The corresponding extrapolated results are presented in bold in column 4 of Table 4  and in columns 4 \& 7 of Tables 5-7 together with some other results including the experimental ones for the low-lying $^{1,3}$S-states wherever available. The term {\it exact} as the subscript of E in the 5th column of Table 5 refers to the results which were obtained by highly accurate variational procedures that involved very large numbers of linear and non-linear parameters [1-2,28-30]. The calculated energies for the low-lying bound $^{1,3}$ states for systems of different nuclear charge Z using data presented in column 3 of Table 4 and columns 3, 4 of Table 5-7  have been plotted in Figure 2 to see how the binding energies of the corresponding states depend on Z. To study the dependence of the convergence trend of calculated energy on Z, the difference in energy $\Delta B=B(N_m+4)-B(N_m)$  is plotted in Figure 3 for few cases of different Z. Using the calculated energy data presented in Table 3 we have plotted Figures 4 and 5 to study the variation of the binding energy B and the quantity $\Delta B$ with increase in $N_m$ for $^{\infty}$Rb$^{84+}$.
The convergence trend in energy can be checked by gradually increasing $N_m$ values in suitable steps and comparing the energy difference $\Delta B=B(N+dN)-B(N)$ with that obtained in the previous step. By analysis of the calculated energy data presented in Table 1, it can be said that the energy of the lowest S-states in the lightest atom under consideration converges much faster than others with respect to the increase in $N_{m}$ values. This trend of convergence with respect to the increasing $N_m$ is slowed down gradually which can be viewed in two ways: (I) with respect to the increase in the level of excitation keeping nuclear charge number Z fixed and (II) with respect to the increase in the nuclear charge number (Z) for a particular level of excitation. For justifying our forgoing remarks -(I) we may compare the convergence trend in energy with respect to $N_m$ for 2$^3$S and 4$^3$S states of neutral helium He (Z=2) or for the corresponding states of highly ionized radon Rn$^{84+}$ (Z=86) by estimating and comparing the energy difference $\Delta B=B(N_m=28)-B(N_m=24)$ using the data presented in columns 4, 6 or those presented in columns 9, 11 of Table 1. These estimates for 2$^3$S and 4$^3$S states of neutral helium (He) are 0.0244au, 0.0660au and those for Rn$^{84+}$ are 31.840au, 117.416au respectively thereby justifying our forgoing remarks -(I).  Similar trends can also be seen in Figure 5 drawn for Rb$^{35+}$ as representative case using data of Table 3. For the justification of our remarks -(II), we may compare the convergence trend in energy with respect to $N_m$ for 4$^1$S state of neutral helium He (Z=2), positively charged rubidium Rb$^{35+}$ (Z=37) and highly charged radon Rn$^{84+}$ (Z=86) by estimating energy difference $\Delta B=B(N_m=28)-B(N_m=24)$ using the data presented in column 5 of Table 1, in column 7 of Table 3 and in column 10 of Table 1 respectively and compare them. And these estimates for 4$^1$S state of He, Rb$^{35+}$, Rn$^{84+}$ are 0.0681au, 19.220au and 102.152au respectively thereby justifying our remarks -(II). Similar results can also be found for remaining levels of excitation using data presented in Table 1-3. This trend of convergence of energy is also demonstrated in Figure 3 for few low-lying $^{1,3}$S states of Rb$^{35+}$ and Rn$^{84+}$ respectively as representative cases. Furthermore it could also be noted that, although, the direct evaluation of the matrix element of  ${\frac{1}{r_{ij}}}$ is possible in the partition $i$ by 
the method of ref. [66], it is not possible for an interaction  other than Coulombic or harmonic oscillator type. For an arbitrary shape of interaction potential, a direct calculation of the matrix element of the potential will involve five dimensional angular integrations which make the calculation very time consuming and leaves door open for inaccuracies to creep in easily. Hence the use of RRC for quick and accurate computation of energy in such cases becomes inevitable.
\begin{figure}
\centering
\fbox{\includegraphics[width=0.95\linewidth, height=0.65\linewidth]{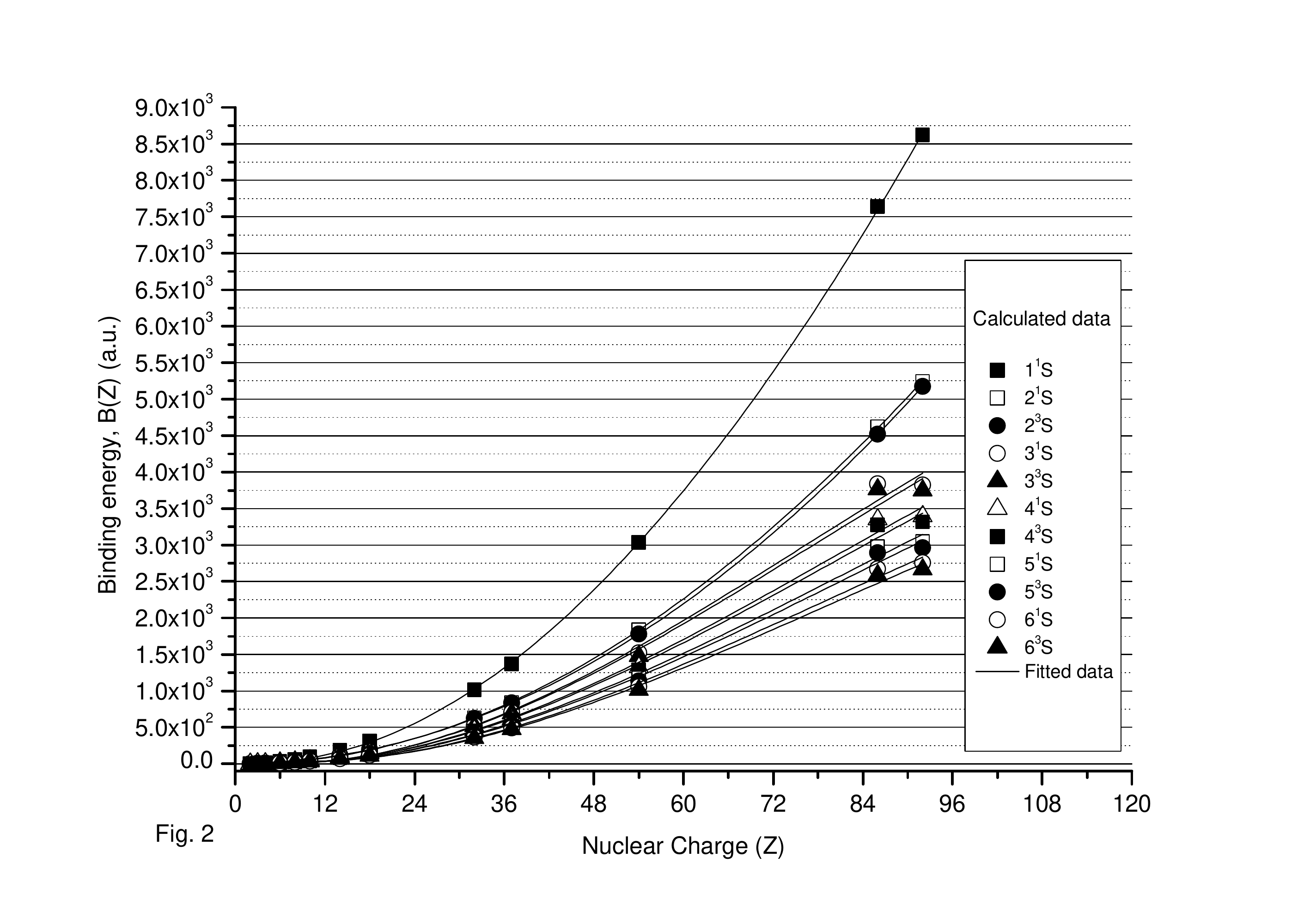}}
\caption{Dependence of the energy, B(=-E) of the low-lying bound $^{1,3}$S-states of helium like Coulombic three-body system $^{\infty}$X$^{(Z-2)+}$ (X=He, Li, Be, C, etc.) on the increase in nuclear charge Z. [Data source: Table 4-7]}
\label{fig:boxed_graphic}
\end{figure}
\newpage
\begin{figure}
\centering
\fbox{\includegraphics[width=0.95\linewidth, height=0.65\linewidth]{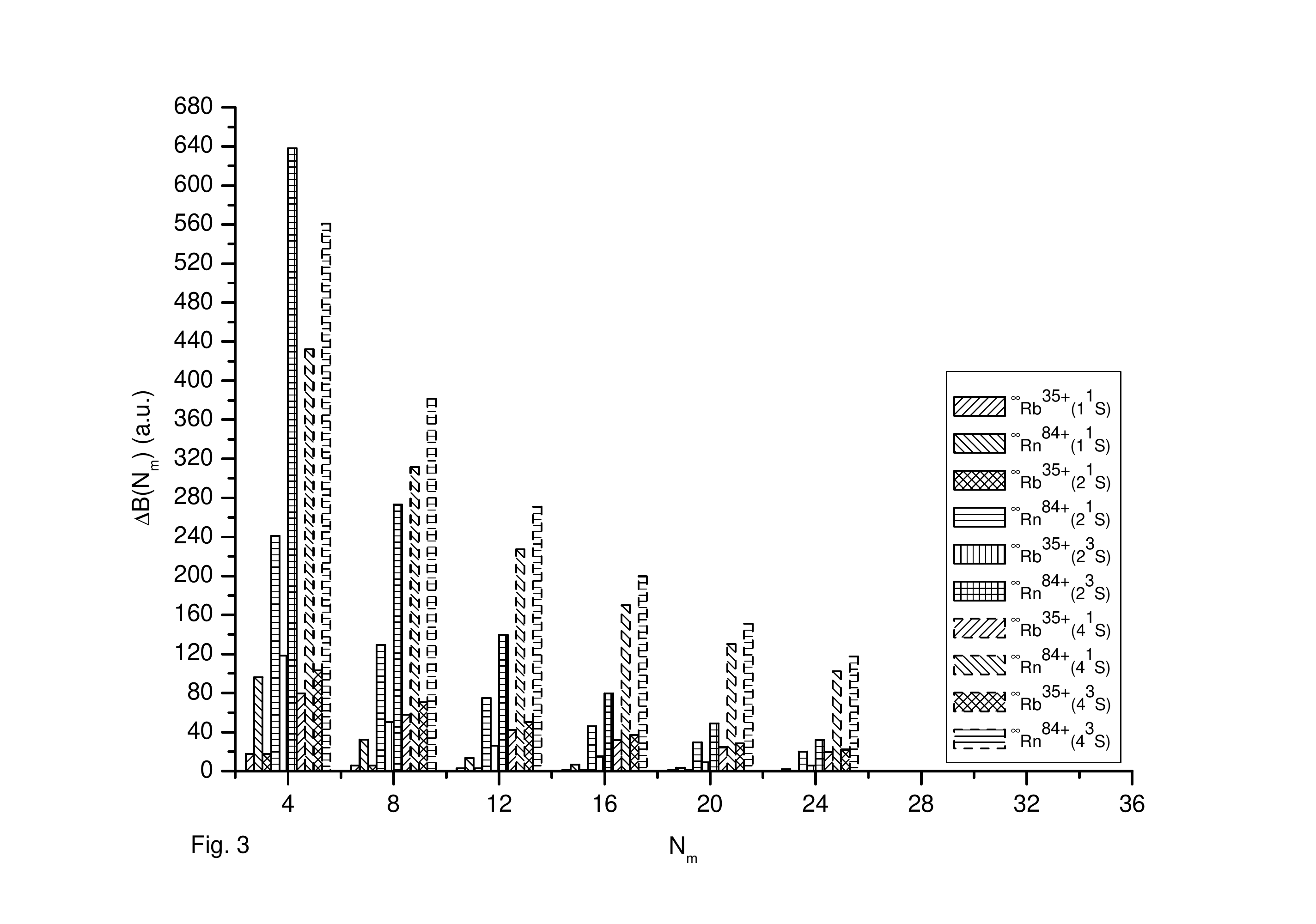}}
\caption{Dependence of the energy difference $\Delta B=B(N_m+4)-B(N_m)$ on the increase in $N_m$ for few low-lying bound $^{1,3}$S-states of helium like Coulombic three-body systems having different nuclear core charge Z. [Data source: Table 1 \& 3]}
\label{fig:boxed_graphic}
\end{figure}
\newpage
\begin{figure}
\centering
\fbox{\includegraphics[width=0.95\linewidth, height=0.65\linewidth]{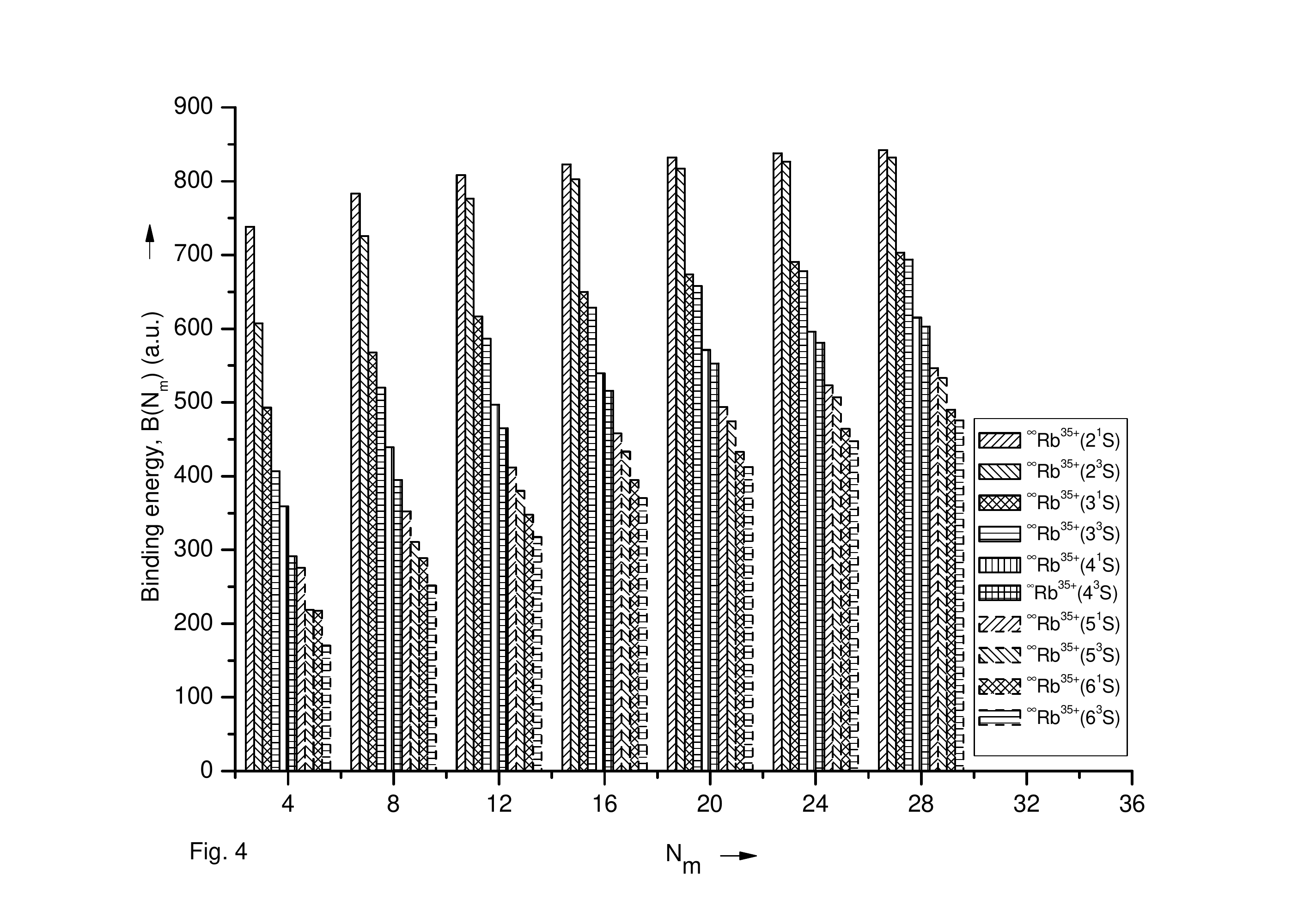}}
\caption{Dependence of the energy B (=-E) on the increase in $N_m$ for few low-lying bound $^{1,3}$S-states of $^{\infty}$Rb$^{35+}$ ion. [Data source: Table 3]}
\label{fig:boxed_graphic}
\end{figure}
\newpage
\begin{figure}
\centering
\fbox{\includegraphics[width=0.95\linewidth, height=0.65\linewidth]{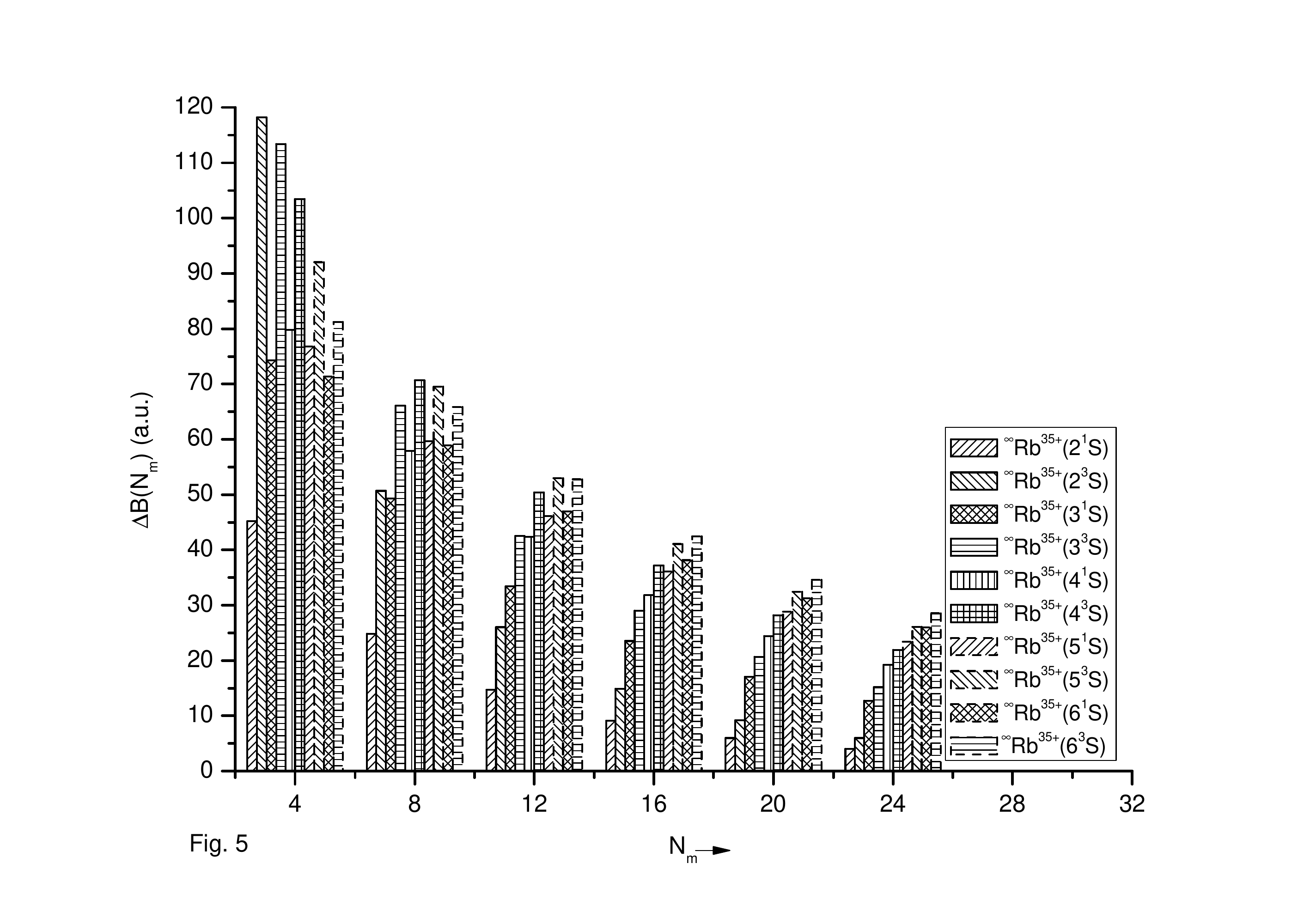}}
\caption{Dependence of the energy difference $\Delta B=B(N_m+4)-B(N_m)$ on the increase in $N_m$ for few low-lying bound $^{1,3}$S-states of $^{\infty}$Rb$^{35+}$ ion.  [Data source: Table 3]}
\label{fig:boxed_graphic}
\end{figure}
One may see from Figure 2, that the calculated energy increases gradually with the increase in the charge number (Z) of the nuclear core of helium like Coulombic three-body systems $^{\infty}$X$^{(Z-2)+}$ (X=He, Li, Be, C,....,U). An approximate value of the energy of the low-lying bound $^{1,3}$S-states of any two-electron three-body system with a given Z can be estimated following empirical formula using the appropriate set of values of parameters $p_0, p_1, p_2, p_3$ recorded in Table 8
\begin{eqnarray}   B(Z)&=& \sum_{j=0}^3p_jZ^j \end{eqnarray}
The values of the parameters $p_0, p_1, p_2, p_3$ presented in Table 8 are obtained by fitting the calculated energy data presented in Table 4 to Table 7 for the low-lying bound $^{1,3}$S-states of two-electron Coulombic three-body systems having nuclear charge (Z) in the range Z=2 to Z=92.

Figure 3 indicates that the rate of convergence in energy with respect to the gradual increase in the size of the basis states $N_m$ in the case of Rn$^{84+}$ (Z=86) is slower than that for Rb$^{35+}$ (Z=37). This convergence trend can be checked by observing the relative decrement in the height of the bars corresponding to a particular level of excitation of Rb$^{35+}$ and  Rn$^{84+}$ and comparing them with respect to gradually increasing values of $N_m$. For example, the height of the extreme left bar representing 1$^1$S state of Rb$^{35+}$ decreases quicker than the height of the bar on its adjacent right representing 1$^1$S state of Rn$^{84+}$ with increasing values of $N_m$. Similar observations hold for other quantum states (or levels of excitation) of the systems.

Further we have plotted in Figures 4 and 5 the variation of binding energy B, and the quantity $\Delta B$ against $N_m$ for different low-lying bound $^{1,3}$S-states of Rb$^{35+}$ ion as a representative case to study the convergence trend of energy of the low-lying bound S-states of a system with fixed Z using the data of Table 3. And it is observed from Figures 4 and 5, that the energies obtained for relatively lower levels converges earlier than the higher ones. Finally in Tables 4-7, the energies of the low-lying bound $^{1,3}$S-states for several Coulombic three-body systems obtained by solving the truncated set of coupled differential equations by the renormalized Numerov method [48] in the framework of HHE aided by RRC, have been compared with the ones of the literature.

\section{Conclusion}
\hspace*{1cm} In conclusion, we note that the use of RRC in HHE method becomes inevitable for the solution of the three-body Schr\H{o}dinger equation, if the inter-particle interaction is other than Coulomb or harmonic type. Hence these coefficients are of immense importance for any type of interaction involved in three-body calculation. However in Tables 4, the calculated energy of the bound $^{1,3}$ S-state at $N_m$=28  in most cases are smaller than those listed in column 5 \& 6. This is due to the eventual truncation of expansion basis to a maximum value of N up to $N_m$=28 due to computer memory limitation. However, one may extrapolate the calculated energy values for $N_m$=0, 4, 8,... etc. to get the solution for still higher $N_m$>28 which have been described in the previous section and the corresponding extrapolated data (for $N_m$>$28$) have been demonstrated in Table 2  for few low-lying S-states of He and Rn as representative cases. The extrapolated energy values (at $N_m$=$N_M$) are listed in bold in the $4^{th}$ column of Table 4 and in columns 4,7 of Tables 5-7. The extrapolated energies agree fairly well with the corresponding exact values found in the literature. One of the important aspect of RRC's are that they are independent of r, and need to be calculated once only and stored, resulting in an economic and highly efficient  numerical computation. Finally we note that by the present method one can describe the Coulombic three-body systems in a very systematic and elegant manner with assured convergence. The method could also be applied to more complex Coulombic or nuclear systems by the proper choice of inter-particle potentials and expansion basis.\\
\hspace*{1cm} The author gladly acknowledges the computational facilities extended by Aliah University for this work.

\newpage
\section{Figure caption}
\vspace{1cm}
\parskip 0.5cm
\parindent 0cm
Figure 1. Particle label scheme for general three-body system and choice of Jacobi coordinates in the $i^{th}$ partition. 

Figure 2. Dependence of the energy, B(=-E) of the low-lying bound $^{1,3}$S-states of helium like Coulombic three-body system $^{\infty}$X$^{(Z-2)+}$ (X=He, Li, Be, C, etc.) on the increase in nuclear charge Z. [Data source: Table 4-7]

Figure 3. Dependence of the energy difference $\Delta B=B(N_m+4)-B(N_m)$ on the increase in $N_m$ for few low-lying bound $^{1,3}$S-states of helium like Coulombic three-body systems having different nuclear core charge Z. [Data source: Table 1 \& 3]

Figure 4. Dependence of the energy B (=-E) on the increase in $N_m$ for few low-lying bound $^{1,3}$S-states of $^{\infty}$Rb$^{35+}$ ion. [Data source: Table 3]

Fig. 5. Dependence of the energy difference $\Delta B=B(N_m+4)-B(N_m)$ on the increase in $N_m$ for few low-lying bound $^{1,3}$S-states of $^{\infty}$Rb$^{35+}$ ion.  [Data source: Table 3]

\section{Tables}
\newpage
\begin{table}
\begin{center}
\begin{scriptsize}
{\bf Table 1. Convergence trend of the energy calculated for few low-lying bound $^{1,3}$S-states of neutral helium ($^{\infty}$He) and highly charged radon ($^{\infty}$Rn$^{84+}$) for increasing $N_m$.}\\
\vspace{5pt}
\begin{tabular}{rllllllllll}\hline
 &\multicolumn{10}{c}{Binding energy (=$B_{N_m}$ for $N=N_m$) in atomic unit (a.u.) in the $^{1,3}S$ state of: }
\\
\cline{2-10}\\
$N_m$&He$^{} _{1^1S}$&He$^{} _{2^1S}$&He$^{} _{2^3S}$&He$^{} _{4^1S}$ & 
He$^{} _{4^3S}$&Rn$^{84+} _{1^1S}$&Rn$^{84+} _{2^1S}$&Rn$^{84+} _{2^3S}$ 
&Rn$^{84+} _{4^1S}$&Rn$^{84+} _{4^3S}$ \\\hline\hline
0 & 2.5000	   &1.2755    &   		&0.5109  &       & 7065.857   &3573.695      &     & 1438.793&\\
4 & 2.7844      &1.5993  	 & 1.3725    &0.8080  &0.6551 &7482.252  & 4076.054      &3309.720  & 1977.548 &1591.887\\
8  &2.8562     & 1.7715  	 & 1.7268  & 1.0414  &0.9524 & 7578.353      & 4316.945      & 3947.931      &2410.031 &2152.732 \\  	
12	&2.8760       & 1.8785  	& 1.8893    &1.2133     &1.1600  & 7610.622      & 4446.135      &4220.691      &2721.692&2534.140\\  
16 	&2.8875     & 1.9477     	 & 1.9788 & 1.3411     &1.3106  &7624.146       & 4520.970      &4360.459          &2948.823&2805.175\\  
 20     &2.8936   &1.9946        & 2.0332   & 1.4378     &1.4227      &7630.732      & 4567.020      & 4440.276           &3118.704 &3004.615\\  
24     & 2.8970   & 2.0275      & 2.0687   & 1.5125    &1.5080     &7634.293      & 4596.738      &4489.347    &3249.001&3155.771\\  
28	 &  2.8990      & 2.0541 	& 2.0931 &1.5706     &1.5740     & 7636.379         & 4616.679     & 4521.187      &3351.153&3273.187\\  
 \\ \hline\hline

\end{tabular}
\end{scriptsize}
\end{center}
\end{table}

\newpage
\begin{table}
\begin{center}
\begin{scriptsize}
{\bf Table 2. Convergence trend of the extrapolated energy obtained for few low-lying bound $^{1,3}$S-states of helium ($^{\infty}$He) and ionized radon ($^{\infty}$Rn$^{84+}$) for increasing $N_m$.}\\
\vspace{5pt}
\begin{tabular}{rllllllllll}\hline
 &\multicolumn{10}{c}{Binding energy (=$B_{N_m}$ for $N=N_m$) in atomic unit (a.u.) in the $^{1,3}S$ state of: }
\\
\cline{2-10}\\
$N_m$&He$^{} _{1^1S}$&He$^{} _{2^1S}$&He$^{} _{2^3S}$&He$^{} _{4^1S}$ & 
He$^{} _{4^3S}$&Rn$^{84+} _{1^1S}$&Rn$^{84+} _{2^1S}$&Rn$^{84+} _{2^3S}$ 
&Rn$^{84+} _{4^1S}$&Rn$^{84+} _{4^3S}$ \\\hline\hline

  32	&      2.9003      & 2.0722    	& 2.1091    &1.6184 &1.6255   & 7637.729    & 4630.917     &4542.088      &3434.018 &3363.887\\  
36  &     2.9011       & 2.0858    & 2.1206  & 1.6571  &1.6665  &7638.621      &4641.212 & 4556.607     &3500.912 &  3434.194 \\  
40   &    2.9017       & 2.0963     &2.1289  & 1.6889  &1.6995 & 7639.235     & 4648.840 & 4567.006      &  3555.503 &3491.030 \\  
44 &      2.9021       & 2.1045    &2.1351  & 1.7152     &1.7264 & 7638.670     &4654.612      	&4574.646     &3600.493 &3537.043    \\  
 48 &      2.90240      & 2.1110    &2.1399  & 1.7371     &1.7485   & 7639.988      & 4659.058      	& 4580.386      &3637.905 &3574.690 \\  
52&       2.9026       & 2.1162  & 2.1436      &1.7556    &1.7668      &7640.224     & 4662.539      	& 4584.781      &3669.268&3605.789 \\  
56   &   2.9028         &2.1205    & 2.1465     & 1.7712  &1.7822     & 7640.405 &4665.303      	& 4588.203      &3695.757 &3631.704\\  
 60	&       2.9029      & 2.1240    &2.1488     & 1.7845     &1.7951  & 7640.545	&4667.526           &4590.906     &3718.283&3653.472\\  
64  &    2.9031       & 2.1267   &2.1507   & 1.7960     &1.8061 & 7640.655     &4669.333          	& 4593.071      &3737.560 &3671.889   \\  
  68 &    2.9031       & 2.1293   &2.1522      & 1.8059 &1.8155 & 7640.742  & 4670.817           	&4594.824      &3754.154&  3687.577 \\  
72 &     2.9032       & 2.1313    & 2.1535   & 1.8144   &1.8236 &7640.813      &4672.048      &4596.259   &3768.515 &3701.022   \\  
 76  &    2.9033       & 2.1330     & 2.1545    &1.8219   &1.8305 & 7640.871     &4673.077           & 4597.445     &3781.006  &3712.612 \\  
 80 &      2.9033      & 2.1345    & 2.1554    & 1.8284  &1.8366 & 7640.918      &4673.944      & 4598.434      &3791.923  &3722.656  \\ 
84  &     2.9034       & 2.1358  & 2.1561   & 1.8342 &1.8419 &7640.958      &4674.680           & 4599.265      &3801.506  &3731.402\\  
88  &   2.9034     & 2.1369   & 2.1568     & 1.8393 &1.8465 &7640.991    & 4675.308          &4599.968     &3809.952  &3739.055 \\  
92   &   2.9034      & 2.1378    &2.1573   & 1.8438 &1.8506  & 7641.019      & 4675.848           & 4600.567      &3817.425  &3745.780  \\   
 96    &  2.9034       & 2.1386  &2.1578   & 1.8478  &1.8542 &7641.043    & 4676.314  & 4601.081      &3824.061&3751.713\\   
100  &    2.9035       & 2.1394     & 2.1582   &1.8513      &1.8575 & 7641.063      &4676.719      & 4601.524      &3829.975  &3756.967 \\   
104 &2.9035            & 2.1400      & 2.1585   &1.8545       & 1.8603  & 7641.081      & 4677.072      &4601.908 &3835.261&3761.638    \\   
108 &2.9035          & 2.1406    & 2.1588    & 1.8574   &1.8629  &7641.096      & 4677.382     &    4602.242        &3840.001&3765.803   \\   
 112& 2.9035           & 2.1411     & 2.1591    & 1.8600  & 1.8651 &7641.109     & 4677.654      & 4602.535         &3844.264 &3769.529  \\   
 116 &  2.9035        & 2.1415    & 2.1593   &1.8623 &1.8672 &7641.121     & 4677.895      & 4602.792        &3848.108  &3772.873 \\   
 120  &   2.9035          & 2.1419 & 2.1596  &1.8644 &1.8690 & 7641.131      & 4678.109     & 4603.019      &3851.584 &3775.882 \\   
 124  &   2.9035          & 2.1423  & 2.1597   & 1.8664  & 1.8707 & 7641.140     &4678.299      & 4603.220      &3854.734 &3778.597   \\   
 128  & 2.9035            & 2.1426      & 2.1599  &1.8681 & 1.8722 & 7641.148     &4678.469     & 4603.399      &3857.596 &3781.054   \\
 132  &    2.9036        & 2.1429        & 2.1601    &1.8697 &1.8736 & 7641.155      & 4678.621     & 4603.558     &3860.202 &3783.282 \\
136  &       2.9036     & 2.1431       &2.1602       & 1.8711 &1.8749  & 7641.161      & 4678.758   & 4603.701     &3862.579 &3785.307\\
140  &    2.9036        & 2.1434           & 2.1603      & 1.8725  &1.8760 &7641.167  & 4678.881      & 4603.829      &3864.754  &3787.152  \\
144  &  2.9036         & 2.1436          &2.1604       & 1.8737   & 1.8770  &7641.172   & 4678.992     & 4603.944      &3866.746 &3788.837    \\
   148 &         2.9036&    2.1438 &   2.1605 & 1.8748 & 1.8780 &  7641.176 & 4679.093 &  4604.048&     3868.574& 3790.378\\      
   152 &         2.9036&    2.1439 &  2.1606 & 1.8758 & 1.8789 &   7641.180 &  4679.184 &     4604.143&     3870.255 & 3791.790\\     
   156 &         2.9036&    2.1441&   2.1607 & 1.8768&1.8797 &   7641.184 &  4679.267 &     4604.228&     3871.804 &3793.087\\    
   160 &         2.9036&    2.1442&    2.1608 & 1.8777 & 1.8804 &   7641.187 &  4679.343 &     4604.305&     3873.233 &3794.281\\     
   164 &         2.9036&    2.1444&  2.1608 & 1.8785 & 1.8811 &    7641.190 & 4679.412 &     4604.376&     3874.553 &3795.380\\     
   168 &         2.9036&    2.1445&  2.1609 & 1.8792 & 1.8817 &    7641.193 & 4679.476 &     4604.441&     3875.775 &3796.395\\     
   172 &         2.9036&    2.1446&   2.1609 & 1.8799 & 1.8823 &    7641.195 &  4679.534 &     4604.500&     3876.908 &3797.334\\   
   176 &         2.9036&    2.1447&   2.1610 & 1.8806 &1.8829 &    7641.198 &  4679.587 &     4604.554&     3877.959 &3798.202\\   
   180 &         2.9036&    2.1448&   2.1610 & 1.8812 & 1.8834 &    7641.200 & 4679.637 &4604.604     &     3878.937 &3799.007\\    
   184 &         2.9036&    2.1449&   2.1611 & 1.8817& 1.8838 &    7641.202 & 4679.682 &     4604.649&     3879.846 & 3799.756\\ 
   188 &         2.9036&    2.1450&   2.1611 & 1.8823 &1.8843 &    7641.203 &  4679.724 &     4604.692&     3880.694 &3800.451\\   
   192 &         2.9036&    2.1451&   2.1611 & 1.8827 &1.8847 &    7641.205 &  4679.763 &     4604.730&     3881.485 &3801.099\\    
   196 &         2.9036&    2.1451&   2.1612 & 1.8832 &1.8851 &    7641.207 & 4679.799 &     4604.766&     3882.223 &3801.702\\   
   200 &         2.9036&    2.1452&   2.1612 & 1.8836 &1.8854 &    7641.208 & 4679.832 &     4604.800&     3882.914 &3802.266\\   
   204 &         2.9036&    2.1453&   2.1613 &  1.8840 &1.8857 &    7641.209 &4679.863 &     4604.831&     3883.561 &3802.792\\     
   208 &         2.9036&    2.1453&   2.1613 & 1.8844 & 1.8860 &    7641.211 & 4679.892 &  4604.859&     3884.167 &3803.284\\      
   212 &         2.9036&    2.1454&   2.1613 & 1.8848 &1.8863 &    7641.212 & 4679.919 &     4604.886&     3884.735 &3803.745\\     
   216 &         2.9036&    2.1454&   2.1613 &  1.8851 &1.8866 &    7641.213 & 4679.944 &     4604.911&     3885.269 &3804.177\\       
   220 &         2.9036&    2.1455&   2.1614 & 1.8854 &1.8869 &    7641.214 & 4679.967 &     4604.934&     3885.771 &3804.583\\ \hline\hline
\end{tabular}
\end{scriptsize}
\end{center}
\end{table}

\begin{table}
\begin{center}
\begin{scriptsize}
{\bf Table 3.  Pattern of convergence of the calculated energy (in atomic unit) for the low-lying bound $^{1,3}$S-states $^{\infty}$Rb$^{35+}$ ion for increasing $N_m$.}\\
\vspace{5pt}
\begin{tabular}{llllllllll}\hline
$N_m$& B(1$^1$S)&B(2$^1$S)&B(2$^3$S)&B(3$^1$S)&B(2$^3$S)&B(4$^1$S)&B(4$^3$S)&B(5$^1$S)& B(5$^3$S)\\\hline
0&	1264.570	&645.919	&	&389.957&		&260.452&		&186.103&		 \\	
4&	1340.648	&738.134	&607.378	&493.011	&406.631	&359.347	&291.160	&275.378	&218.704	\\
8&	1358.438	&783.281&	725.557&	567.303&	520.017	&439.137&	394.568&	352.145	&310.756\\
12	&1364.548&	808.142&	776.213&	616.614&	586.098&	497.042&	465.246&	411.821	&380.261\\
16	&1367.139	&822.797	&802.266&	650.037&	628.633&	539.433&	515.660&	457.947&	433.232\\
20&	1368.412	&831.933	&817.185&	673.538&	657.660&	571.245	&552.834&	494.105	&474.305\\
24&	1369.104	&837.886 &826.380&	690.615 &678.330	&595.707 &581.043&	522.943	&506.763	\\
28	&1369.511&	841.914&	832.357	&703.368&	693.539&	614.927	&602.972&	546.324&	532.871\\\hline
\end{tabular}
\end{scriptsize}
\end{center}
\end{table}

\begin{table}
\begin{center}
\begin{scriptsize}
{\bf Table 4. Comparison of calculated energy for the low-lying bound $^{1,3}$S(L=0)-states of helium and helium like Coluombic three-body systems with the ones of the  literature.}\\
\vspace{5pt}
\begin{tabular}{l l l l cc}\hline
System &State&\multicolumn{3}{c}{Binding energy, B (=$B_{N_m}$ for $N=N_m$) in atomic unit (a.u.)}
\\
\cline{3-6}
&& $B_{N_m=28}$ & $B_{N_m=N_M}$ & 
$E_{exact}$ & Other sources \\\hline
$^{3}$He& $1^1S$& 2.89845295&{\bf 2.90324338}& 2.90316721 [35]&2.90051530 [1]\\ 
& $2^1S$& 2.05107241&{\bf 2.13814412}& 2.14558192 [35]&2.14501773 [1]\\ 
&$2^3S$& 2.09269897&{\bf 2.16138053 }&  2.17483231[35]&2.17454273 [1]\\ 
& $3^1S$& 1.78731919&{\bf 2.00852819}&2.06089652 [35]&2.06069722 [1]\\ 
& $3^3S$& 1.80010572&{\bf 2.00295712}&2.06831238 [35]&2.06820440[1]\\ 
& $4^1S$& 1.57029931&{\bf 1.88975703}& 2.03321657 [35]&2.02623837 [1]\\ 
& $4^3S$& 1.57365954&{\bf 1.89174046}& 2.03614146 [35]&2.03156807[1]\\ 

$^{4}$He&$1^1S$& 2.89859016&{\bf 2.90338079}& 2.90330456 [35]&2.90065336 [1]\\
&&&&2.90372440 [24]&2.90368830 [14]\\
&$2^1S$& 2.05116527&{\bf 2.13824141}& 2.14567859 [35]&2.14511445 [1]\\
&$2^3S$& 2.09279342&{\bf 2.16147778}& 2.17493019[35]&2.17464057[1]\\
&$3^1S$& 1.78740060&{\bf 2.00862130}& 2.06098908 [35]&2.06078978 [1]\\
&$3^3S$& 1.80018701&{\bf 2.00304736}& 2.06840524[35]&2.06829724 [1]\\
&$4^1S$& 1.57037897&{\bf 1.88986615}&2.03330782 [35]&2.02633010 [1]\\
&$4^3S$& 1.57373604&{\bf 1.89184097}&2.03623283 [35]&2.03165943[1]\\

$^{\infty}$He&$1^1S$&2.89900954&{\bf 2.90380076}&2.90372438 [35] &2.90107544 [1]\\
&&&&2.90372438 [22] &\\
&$2^1S$&2.05414493&{\bf 2.14642285}& 2.14597405 [35] &2.14541020 [1]\\
&$2^3S$&2.09308211&{\bf 2.16177503}& 2.17522938 [35] &2.17493966 [1]\\
&$3^1S$&1.78764955&{\bf 2.01000443}& 2.06127199 [35] &2.06107284 [1]\\
&$3^3S$&1.80043551&{\bf 2.00332324}& 2.06868907 [35] &2.06858102 [1]\\
&$4^1S$&1.57062240&{\bf 1.89199576}& 2.03358672 [35] &2.02660791 [1]\\
&$4^3S$&1.57396988&{\bf 1.89214821}& 2.03651208 [35] &2.03193872 [1]\\\\

$^{6}$Li$^{+}$&$1^1S$&7.27068287&{\bf 7.27959321}& 7.27922302 [35]&7.27588119 [1]\\
&$2^1S$ &4.88519501 &{\bf 5.03990504} & - &5.03894691 [1]\\
&$2^3S$&4.96035893&{\bf 5.09413803}& -&5.10969189 [1]\\
&$3^1S$&4.21338221&{\bf 4.67184315}& -&4.73266048 [1]\\
&$3^3S$&4.22271122 &{\bf 4.64640852}& -&4.75130451 [1]\\
&$4^1S$&3.72411106&{\bf 4.48575388}& -&4.62472771 [1]\\
&$4^3S$&3.70030123&{\bf 4.43114605}& -&4.63390581 [1]\\
&$5^1S$&3.32869817&{\bf 4.39001212}& -&-\\
&$5^3S$&3.28429267&{\bf 4.31111429}& -&-\\\\

$^{7}$Li$^{+}$&$1^1S$&7.27078128 &{\bf 7.27969173}&7.27932152 [35]&7.20603060 [1]\\
&$2^1S$&4.88525904 &{\bf 5.03997129}&-&5.03913196 [1]\\
&$2^3S$&4.96042382 &{\bf 5.09420455}&-&5.10975862 [1]\\
&$3^1S$&4.21343739 &{\bf4.67190463}&-&4.73315062 [1]\\
&$3^3S$&4.22276636 &{\bf 4.64646894}&-&4.75136656 [1]\\
&$4^1S$&3.724159 &{\bf 4.48581310}&-&4.62641756 [1]\\
&$4^3S$&3.70034950&{\bf 4.43120337}&-&4.63396662 [1]\\
&$5^1S$&3.328742 &{\bf 4.39006911}&-&-\\
&$5^3S$&3.28433555&{\bf 4.31117021}&-&-\\\\

$^{\infty}$Li$^{+}$&$1^1S$&7.27137265 &{\bf 7.28028376}& 7.27991341 [35]&7.27657671 [1]\\ 
&& && 7.27991341 [22]&7.27991341[74]\\ 
&$2^1S$&4.88564379 &{\bf 5.04036937}& 5.04087674 [73]&5.03941025 [1]\\
&$2^3S$&4.96081374 &{\bf 5.09460425}& 5.11072731 [73]&5.11015939 [1]\\
&&&& 5.11072737 [20]&\\
&$3^1S$&4.21376895 &{\bf 4.67227404}&4.73375186 [73]&4.73309441 [1]\\
&$3^3S$&4.22309769 &{\bf 4.64683203}&4.75207644 [73]&4.75173831 [1]\\
&$4^1S$&3.72445246 &{\bf 4.48616767}&4.62977459 [73]&4.62515508 [1]\\
&$4^3S$&3.70063956 &{\bf 4.43154788}&4.63713654 [73]&4.63433035 [1]\\
&$5^1S$&3.32900485 &{\bf 4.39042199}&-&-\\
&$5^3S$&3.28459324 &{\bf 4.31150622}&-&-\\\hline
\end{tabular}
\end{scriptsize}
\end{center}
\end{table}

\begin{table}
\begin{center}
\begin{scriptsize}
{\bf Table 5.  Comparison of calculated energy for the low-lying bound $^{1,3}$S(L=0)-states of helium like Coluombic three-body systems with the ones of the  literature.}\\
\vspace{5pt}
\begin{tabular}{l l r l l r l}\hline
&\multicolumn{6}{c}{Binding energies $B_{N_m}$ for $N=N_m$ in atomic unit for $^{1,3}$S- states}\\
\cline{2-7}
System&State& $B_{N_m=28}$ & $B_{N_m=N_M}$ &State& $B_{N_m=28}$ & $B_{N_m=N_M}$ \\\hline

$^{10}$Be$^{2+}$&$1^1S$&13.64100542 &{\bf 13.65555317}&&&\\
&&&13.6555662 [24]&&&\\
&&&13.6555322 [14]&&&\\
&$2^1S$&8.94911216 &{\bf 9.19161005}&$2^3S$&9.05459257 &{\bf 9.27634522}\\
&$3^1S$&7.66804423 &{\bf 8.44652588}&$3^3S$&7.66497104  &{\bf 8.38910901}\\
&$4^1S$&6.76103019 &{\bf 8.08313611}&$4^3S$&6.70286146 &{\bf 7.97128862}\\

$^{\infty}$Be$^{2+}$& $1^1S$&13.64177142 &{\bf 13.65631993}&&&\\
&&&13.65566238[22]&&&\\
&&&13.65566238[74]&&&\\ 
& $2^1S$&8.94960223 &{\bf 9.19333838}& $2^3S$&9.27685193 &{\bf 9.27685193}\\
&&&&&&9.29716659 [20] \\
& $3^1S$&7.66846380 &{\bf 8.45124727}& $3^3S$&7.66538960 &{\bf 8.38956578} \\
& $4^1S$&6.76066022&{\bf 8.08949853}& $4^3S$&6.70322721 &{\bf 7.97172093} \\
& $5^1S$&6.03494406 &{\bf 7.90747336}& $5^3S$&5.94304043&{\bf 7.74863876} \\\\

$^{12}$C$^{4+}$&$1^1S$&32.37663926 &{\bf 32.40685560}&&&\\
&&&32.4062466 [24]&&&\\
&&&32.4062132 [14]&&&\\
&$2^1S$&20.77005736 &{\bf 21.24818024}&$2^3S$&20.92259034 &{\bf 21.38852080}\\
&$3^1S$&17.66499856 &{\bf 19.33438639}&$3^3S$&17.60913254 &{\bf 19.17442075}\\
&$4^1S$&15.53267060 &{\bf 18.43106051}&$4^3S$&15.36613898 &{\bf 18.15220701}\\
&$5^1S$&13.84303207 &{\bf 17.96048254}&$5^3S$&13.60727634 &{\bf 17.59965248}\\

$^{\infty}$C$^{4+}$&$1^1S$&32.37814008 &{\bf 32.40835778}&&&\\
&&&32.40624660[22 ]&&&\\
&&&32.40624660[73]&&& \\
&$2^1S$&20.77100359 &{\bf 21.25084519}&$2^3S$&20.92354286 &{\bf 21.38949376}\\
&&&&&&21.4207559 [20] \\
&$3^1S$&17.66580282 &{\bf 19.34319951}&$3^3S$&17.60993325 &{\bf 19.17529100}\\
&$4^1S$&15.53337831 &{\bf 18.44829391}&$4^3S$&15.36683732&{\bf 18.15302868}\\
&$5^1S$&13.84366250 &{\bf 17.98746303}&$5^3S$&13.60789448 &{\bf 17.60044649}\\
&$6^1S$&12.44587479 &{\bf 17.78810696}&$6^3S$&12.16585900 &{\bf 14.83317733}\\\\

$^{16}$O$^{6+}$&$1^1S$&59.10834846&{\bf 59.16004356}&&&\\
&&&59.156 5951 [24]&&&\\
&&&59.1565622 [14]&&&\\
&$2^1S$&37.51773633&{\bf 38.31230986}&$2^3S$&37.69735366&{\bf 38.49829459}\\
&$3^1S$&31.77962231&{\bf 34.67519168}&$3^3S$&31.63349495&{\bf 34.36036915}\\
&$4^1S$&27.90383672&{\bf 32.99044656}&$4^3S$&27.57415238&{\bf 32.46790039}\\
&$5^1S$&24.84856668&{\bf 32.10636657}&$5^3S$&24.40313985&{\bf 31.44020700}\\
&$6^1S$&22.32748157&{\bf 31.59904473}&$6^3S$&21.80904495 &{\bf 30.92526993}\\

$^{\infty}$O$^{6+}$&$1^1S$&59.11039341 &{\bf 59.16209025}&&&\\
&&& 59.15659512 [22] &&&\\
&&&59.15659512 [73]&&& \\
&$2^1S$&37.51901718 &{\bf 38.31361890}&$2^3S$&37.69863997 &{\bf 38.49960746}\\
&&&&&&38.54464732 [20]\\
&$3^1S$&31.78070668 &{\bf 34.67637674}&$3^3S$&31.63457330 &{\bf 34.36153886}\\
&$4^1S$&27.90478889 &{\bf 32.99157418}&$4^3S$&27.57509196 &{\bf 32.46900346} \\
&$5^1S$&24.84941473 &{\bf 32.10746357}&$5^3S$&24.40397111&{\bf 31.44127245}\\
&$6^1S$&22.32824372 &{\bf 31.60009809}&$6^3S$&21.80904495&{\bf 30.98033807} \\

$^{20}$Ne$^{8+}$&$1^1S$&93.83962679&{\bf 93.92037567}&&&\\
&&&93.9068065 [24]&&&\\
&&&93.9067737 [14]&&&\\
&$2^1S$&59.19300213&{\bf 60.38347654}&$2^3S$&59.37879471&{\bf 60.60552089}\\
&$3^1S$&50.01242982&{\bf 54.46894570}&$3^3S$&49.73801357&{\bf 53.94690740}\\
&$4^1S$&43.87450984&{\bf 51.75982348}&$4^3S$&43.32683791&{\bf 50.91827229}\\
&$5^1S$&39.05150768&{\bf 50.33364370}&$5^3S$&38.33024314&{\bf 49.26976804}\\

$^{\infty}$Ne$^{8+}$&$1^1S$&93.84221663&{\bf 93.92117897}&&&\\
&&&93.90680652 [22]&&&\\
&&& 93.90680651 [73] &&&\\
&$2^1S$&59.19521811&{\bf 60.38674068}&$2^3S$&59.38041492&{\bf 60.60717382}\\
&&&&&&60.66864658 [20]\\
&$3^1S$&50.01379450&{\bf 54.46665522}&$3^3S$&50.01279450&{\bf 53.94837663}\\
&$4^1S$&43.87570737&{\bf 51.76123767}&$4^3S$&43.32801878&{\bf 50.91965680}\\
&$5^1S$&39.05257343&{\bf 50.33501851}&$5^3S$&38.33128755&{\bf 49.27110500} \\
&$6^1S$&35.07883941&{\bf 49.78428221}&$6^3S$&34.24708372&{\bf 48.51818062}\\\hline
\end{tabular}
\end{scriptsize}
\end{center}
\end{table}

\begin{table}
\begin{center}
\begin{scriptsize}
{\bf Table 6. Calculated energy for the low-lying bound $^{1,3}$S(L=0)-states of helium like Coluombic three-body systems for which reference values are not available.}\\
\vspace{5pt}
\begin{tabular}{l l r r l r r}\hline
&\multicolumn{6}{c}{Binding energies $B_{N_m}$ for $N=N_m$ in atomic unit for $^
{1,3}$S- states}\\
\cline{2-7}
System&State& $B_{N_m=28}$ & $B_{N_m=N_M}$ &State& $B_{N_m=28}$ & $B_{N_m=N_M}$ \\\hline

$^{28}$Si$^{12+}$&$1^1S$&187.33623586&{\bf 187.48697007 }&&&\\
&$2^1S$&117.34098803&{\bf 119.56568139}&$2^3S$&117.46178209&{\bf 119.81233419}\\
&$3^1S$&98.83846617&{\bf 107.41844030}&$3^3S$&98.18763000&{\bf 106.32183570}\\
&$4^1S$&86.61882068&{\bf 101.92547321}&$4^3S$&85.46630711&{\bf 100.22303947}\\
&$5^1S$&77.05326349&{\bf 99.03169898}&$5^3S$&75.57824630&{\bf  96.89583100}\\

$^{\infty}$Si$^{12+}$&$1^1S$&187.33991986&{\bf 187.49065697}&&&\\
&$2^1S$&117.34327599&{\bf 119.56801338}&$2^3S$&117.46407024 &{\bf 119.81466738}\\
&$3^1S$&98.84039233&{\bf 107.42053496}&$3^3S$& 98.18954155&{\bf 106.32390404}\\
&$4^1S$&86.62050862&{\bf 101.93205784}&$4^3S$& 85.46797059&{\bf 100.22498691}\\
&$5^1S$&77.05476512&{\bf  99.03362906}&$5^3S$& 75.57971705&{\bf 96.89771104}\\
&$6^1S$&69.18781295&{\bf 97.87364713}&$6^3S$& 67.50779834&{\bf 95.34894004}\\\\

$^{40}$Ar$^{16+}$&$1^1S$&313.01347510&{\bf 313.26016489}&&&\\
&$2^1S$&195.25738776&{\bf 198.82695786}&$2^3S$&195.17203766&{\bf 199.00936113}\\
&$3^1S$&164.16773363&{\bf 178.19631514}&$3^3S$&162.95847989&{\bf 176.29956177}\\
&$4^1S$&143.78433327&{\bf 168.94885492}&$4^3S$&141.78501941&{\bf 166.06665428}\\
&$5^1S$&127.86367569&{\bf 164.04660012}&$5^3S$&125.35175859&{\bf 160.47777379}\\

$^{\infty}$Ar$^{16+}$&$1^1S$&313.01778694&{\bf 313.26290079}&&&\\
&$2^1S$&195.26005541&{\bf 198.82967402}&$2^3S$&195.17469823&{\bf  199.00308521}\\
&$3^1S$&164.16997440&{\bf 178.19874736}&$3^3S$& 162.96070028&{\bf 176.25561505}\\
&$4^1S$&143.78629491&{\bf 168.95116040}&$4^3S$&141.78695094&{\bf 165.96191861}\\
&$5^1S$&127.86542065&{\bf  164.04954287}&$5^3S$&125.35346601&{\bf 160.29640320}\\
&$6^1S$&114.78563676&{\bf 162.06610437}&$6^3S$&111.94830964&{\bf 157.57856468}\\\\

$^{73}$Ge$^{30+}$&$1^1S$&1014.10495117&{\bf 1014.83717742}&&&\\
&$2^1S$&626.14203646&{\bf 636.44668213}&$2^3S$&621.72918896&{\bf 633.59828623}\\
&$3^1S$& 524.12494015&{\bf 567.05022289}&$3^3S$&518.19512337&{\bf 559.69832511}\\
&$4^1S$&458.44873845&{\bf 536.5722955}&$4^3S$&450.57357075&{\bf 526.41874595}\\
&$5^1S$&407.40409820&{\bf 520.37147652}&$5^3S$&398.21122657&{\bf 508.07162153}\\
&$6^1S$&365.55993388&{\bf 513.54614703}&$6^3S$&355.54564542&{\bf 499.14527714}\\

$^{\infty}$Ge$^{30+}$&$1^1S$&1014.11284946&{\bf 1014.84667354}&&&\\
&$2^1S$&626.14680929&{\bf 636.48612374}&$2^3S$&621.73383114&{\bf 633.63074465}\\
&$3^1S$&524.12890560&{\bf 567.25338865}&$3^3S$&518.19899158&{\bf 559.84659169}\\
&$4^1S$&458.45220105&{\bf 537.01104413}&$4^3S$&450.57693397&{\bf 526.75737901}\\
&$5^1S$&407.40717276&{\bf 521.09516457}&$5^3S$&398.21419883&{\bf 508.65112038}\\
&$6^1S$&365.56269120&{\bf 514.59886056}&$6^3S$&355.54829917&{\bf  500.00969512}\\\\

$^{87}$Rb$^{35+}$&$1^1S$&1369.50177024&{\bf 1370.45052824}&&&\\
&$2^1S$&841.90857190&{\bf 855.03234476}&$2^3S$&832.35208572&{\bf 848.16473859}\\
&$3^1S$&703.36387502&{\bf 759.78528149}&$3^3S$&693.53427843&{\bf 748.88959756}\\
&$4^1S$&614.92316566&{\bf 718.35308809}&$4^3S$&602.96829556&{\bf 704.21342083}\\
&$5^1S$&546.32050458&{\bf 696.26001256}&$5^3S$&532.86744847&{\bf 679.56972477}\\
&$6^1S$&490.12802196&{\bf 647.66171793}&$6^3S$&475.75927472&{\bf 667.54281907}\\

$^{\infty}$Rb$^{35+}$&$1^1S$&1369.51091586&{\bf 1370.46173546}&&&\\
&$2^1S$&841.91402943&{\bf 855.08151199}&$2^3S$&832.357300378&{\bf 848.20695939}\\
&$3^1S$&703.36837784&{\bf 760.05044604}&$3^3S$&693.53862264&{\bf 749.08631627}\\
&$4^1S$&614.92709123&{\bf 718.93145490}&$4^3S$&602.97207243&{\bf 704.66435709}\\
&$5^1S$&546.32398759&{\bf 696.26438420}&$5^3S$&532.87078631&{\bf 680.34234635}\\
&$6^1S$&490.13114408&{\bf 647.66577131}&$6^3S$&475.76225492&{\bf 667.54698910}\\\\

$^{132}$Xe$^{52+}$&$1^1S$&3034.33032264&{\bf 3036.10652657}&&&\\
&$2^1S$&1835.73342239&{\bf 1858.27421167}&$2^3S$&1778.07908319&{\bf 1811.55917013}\\
&$3^1S$&1520.10383171&{\bf1631.61384634}&$3^3S$&1480.70025535&{\bf 1598.25546168}\\
&$4^1S$&1326.34233956&{\bf 1537.17817352}&$4^3S$&1287.14195751&{\bf 1502.67978553}\\
&$5^1S$&1177.30195319&{\bf 1486.00908806}&$5^3S$&1137.44912969&{\bf 1450.16494641}\\
&$6^1S$&1055.62293041&{\bf 1432.37045117}&$6^3S$&1015.54664869&{\bf 1424.69625078}\\
&$7^1S$&953.54597837&{\bf 1365.70263779}&$7^3S$&913.71200484&{\bf 1420.79340749}\\

$^{\infty}$Xe$^{52+}$&$1^1S$&3034.34403549&{\bf 3036.12024452}&&&\\
&$2^1S$&1835.74140513&{\bf1858.28223928}&$2^3S$&1778.08642782&{\bf 1811.56665120}\\
&$3^1S$&1520.11029918&{\bf 1631.62069230}&$3^3S$&1480.70637251&{\bf 1598.26205731}\\
&$4^1S$&1326.34795821&{\bf 1537.18457058}&$4^3S$&1287.14727680&{\bf 1502.68598043}\\
&$5^1S$&1177.30693158&{\bf 1486.01523381}&$5^3S$&1137.45383214&{\bf 1450.17091643}\\
&$6^1S$&1055.62738998&{\bf 1432.37654976}&$6^3S$&1015.55084882&{\bf 1424.70210581}\\
&$7^1S$&953.55000451 &{\bf 1365.70825884}&$7^3S$&913.71578525&{\bf 1420.79923485}\\\hline
\end{tabular}
\end{scriptsize}
\end{center}
\end{table}

\begin{table}
\begin{center}
\begin{scriptsize}
{\bf Table 7. Calculated energy for the low-lying bound $^{1,3}$S(L=0)-states of helium like Coluombic three-body systems for which reference values are not available.}\\
\vspace{5pt}
\begin{tabular}{l l r r l r r}\hline
&\multicolumn{6}{c}{Binding energies $B_{N_m}$ for $N=N_m$ in atomic unit for $^
{1,3}$S- states}\\
\cline{2-7}
System&State& $B_{N_m=28}$ & $B_{N_m=N_M}$ &State& $B_{N_m=28}$ & $B_{N_m=N_M}$ \\\hline

$^{222}$Rn$^{84+}$&$1^1S$&7636.36418198&{\bf 7641.20543961}&&&\\
&$2^1S$&4616.66857975&{\bf 4680.19668342}&$2^3S$&4521.17553625&{\bf 4605.15844138}\\
&$3^1S$&3837.15245292&{\bf 4136.94140731}&$3^3S$&3764.45017663&{\bf 4058.24848136}\\
&$4^1S$&3351.14526034&{\bf 3893.40869622}&$4^3S$&3273.17906922&{\bf 3810.56957110}\\
&$5^1S$&2976.42342407&{\bf 3765.55061894}&$5^3S$&2893.66454225&{\bf 3671.16419665}\\
&$6^1S$&2670.29151481&{\bf 3593.02119421}&$6^3S$&2584.74356269&{\bf 3599.15482185}\\\\

$^{\infty}$Rn$^{84+}$&$1^1S$&7636.37941153&{\bf 7641.23108390}&&& \\
&$2^1S$&4616.67861400&{\bf 4680.41269088}&$2^3S$&4521.18661942&{\bf 4605.36483601}\\
&$3^1S$&3837.16123452&{\bf 4138.31425610}&$3^3S$& 3764.45941573&{\bf 4059.27113623}\\
&$4^1S$&3351.15302320&{\bf 3896.39350957}&$4^3S$&3273.18711492 &{\bf 3812.93142451}\\
&$5^1S$&2976.43036542&{\bf 3767.26785710}&$5^3S$&2893.67166846&{\bf 3675.21000221}\\
&$6^1S$&2670.29777418&{\bf 3599.22653486}&$6^3S$&2584.74994153&{\bf 3605.18235959}\\\\

$^{238}$U$^{90+}$&$1^1S$&8618.00391229&{\bf 8624.11550275}&&&\\
&$2^1S$&5240.84978695 &{\bf 5319.84869467}&$2^3S$&5174.74362888&{\bf 5270.93881700}\\
&$3^1S$&3819.79967898 &{\bf 4449.03430062}&$3^3S$&3746.84431623&{\bf 4360.91561578}\\
&$4^1S$&3394.24156455 &{\bf 4305.79766260}&$4^3S$&3312.84455557&{\bf 4200.53296546}\\
&$5^1S$&3046.23137764 &{\bf 3745.06335941}&$5^3S$&2959.62690696&{\bf 4182.50227889}\\
&$6^1S$&2753.99116703 &{\bf 3988.35675960}&$6^3S$&2664.58714689&{\bf 3581.10905621}\\\\

$^{\infty}$U$^{90+}$&$1^1S$&8618.01897182&{\bf 8624.13058963}&&&\\
&$2^1S$&5240.86018734&{\bf 5319.85951984}&$2^3S$&5174.75544789&{\bf 5271.17435433}\\
&$3^1S$&3819.80783330&{\bf 4449.04422383}&$3^3S$&3746.85290227&{\bf 4360.92558161}\\
&$4^1S$&3394.24887082&{\bf 4305.80737378}&$4^3S$&3312.85216467&{\bf 4205.14588567}\\
&$5^1S$&3046.23797736&{\bf 3745.07101005}&$5^3S$&2959.63372271&{\bf 4190.22906381}\\
&$6^1S$&2753.99716714&{\bf 3988.36584661}&$6^3S$&2664.59330043&{\bf 3581.11748487}\\\hline

\end{tabular}
\end{scriptsize}
\end{center}
\end{table}

\newpage
\begin{table}
\begin{center}
\begin{scriptsize}
{\bf Table 8. Values of parameters involved in eq.(26) obtained by best fit of calculated energies.}\\
\vspace{5pt}
\begin{tabular}{lcccc}\hline
State&\multicolumn{4}{c}{Parameters (in atomic unit, 1au=27.12eV)}
\\
\cline{2-5}
&$p_0$ & $p_1$ & $p_2$ &$p_3$ \\\hline
1$^1$S&28.8078&-7.33198&1.26212&-0.00178\\
2$^1$S&10.37378&-2.67664&0.71274&-7.02147E-4\\
2$^3$S&0.18499&-0.2216&0.61394&-1.53093E-6\\
3$^1$S&63.82991&-14.21409&1.04226&-0.00462\\
3$^3$S&59.93207&-13.24075&0.99848&-0.00435\\
4$^1$S&49.16737&-10.94696&0.85611&-0.00355\\
4$^3$S&46.22096&-10.21724&0.8198&-0.00335\\
5$^1$S&53.3078&-10.35107&0.76665&-0.00314\\
5$^3$S&49.91539&-9.62781&0.7305&-0.00294\\
6$^1$S&87.92359&-12.89187&0.76764&-0.0033\\
6$^3$S&82.89113&-11.97116&0.72684&-0.00307\\\hline
\hline
\end{tabular}
\end{scriptsize}
\end{center}
\end{table}

\end{document}